\pdfoutput=1

\documentclass[12pt,a4paper]{article}

\usepackage{ifthen} 
\newboolean{pdflatex}
\setboolean{pdflatex}{true} 

\newboolean{articletitles}
\setboolean{articletitles}{true} 

\newboolean{uprightparticles}
\setboolean{uprightparticles}{false} 

\newboolean{inbibliography}
\setboolean{inbibliography}{false} 

\usepackage{microtype}
\usepackage{lineno}  
\usepackage{xspace} 
\usepackage{caption} 

\usepackage{graphicx}  
\usepackage{color}
\usepackage{colortbl}
\graphicspath{{./figs/}} 

\usepackage{amsmath} 
\usepackage{amssymb}
\usepackage{amsfonts}
\usepackage{upgreek} 

\newcommand*\patchAmsMathEnvironmentForLineno[1]{%
\expandafter\let\csname old#1\expandafter\endcsname\csname #1\endcsname
\expandafter\let\csname oldend#1\expandafter\endcsname\csname
end#1\endcsname
 \renewenvironment{#1}%
   {\linenomath\csname old#1\endcsname}%
   {\csname oldend#1\endcsname\endlinenomath}%
}
\newcommand*\patchBothAmsMathEnvironmentsForLineno[1]{%
  \patchAmsMathEnvironmentForLineno{#1}%
  \patchAmsMathEnvironmentForLineno{#1*}%
}
\AtBeginDocument{%
\patchBothAmsMathEnvironmentsForLineno{equation}%
\patchBothAmsMathEnvironmentsForLineno{align}%
\patchBothAmsMathEnvironmentsForLineno{flalign}%
\patchBothAmsMathEnvironmentsForLineno{alignat}%
\patchBothAmsMathEnvironmentsForLineno{gather}%
\patchBothAmsMathEnvironmentsForLineno{multline}%
\patchBothAmsMathEnvironmentsForLineno{eqnarray}%
}

\usepackage{hyperref}    
\usepackage[all]{hypcap} 

\def\lhcb {\mbox{LHCb}\xspace}

\def\MagUp {\mbox{\em Mag\kern -0.05em Up}\xspace}

\ifthenelse{\boolean{uprightparticles}}%
{

 \def\Ppi         {\ensuremath{\uppi}\xspace}

 \def\PDelta      {\ensuremath{\Delta}\xspace}                 
 \def\PXi      {\ensuremath{\Xi}\xspace}                 
 \def\PLambda      {\ensuremath{\Lambda}\xspace}                 
 \def\PSigma      {\ensuremath{\Sigma}\xspace}                 
 \def\POmega      {\ensuremath{\Omega}\xspace}                 
 \def\PUpsilon      {\ensuremath{\Upsilon}\xspace}                 
 
 \def\PB      {\ensuremath{\mathrm{B}}\xspace}                 
                  
 \def\PD      {\ensuremath{\mathrm{D}}\xspace}

 \def\PK      {\ensuremath{\mathrm{K}}\xspace}

 \def\Pb      {\ensuremath{\mathrm{b}}\xspace}                 
 \def\Pc      {\ensuremath{\mathrm{c}}\xspace}

 \def\Pi      {\ensuremath{\mathrm{i}}\xspace}

 \def\Pp      {\ensuremath{\mathrm{p}}\xspace}

 \def\Ps      {\ensuremath{\mathrm{s}}\xspace}

}
{

 \def\Ppi         {\ensuremath{\pi}\xspace}

 \mathchardef\PDelta="7101
 \mathchardef\PXi="7104
 \mathchardef\PLambda="7103
 \mathchardef\PSigma="7106
 \mathchardef\POmega="710A
 \mathchardef\PUpsilon="7107
                  
 \def\PB      {\ensuremath{B}\xspace}                 
                  
 \def\PD      {\ensuremath{D}\xspace}

 \def\PK      {\ensuremath{K}\xspace}

 \def\Pb      {\ensuremath{b}\xspace}                 
 \def\Pc      {\ensuremath{c}\xspace}

 \def\Pi      {\ensuremath{i}\xspace}

 \def\Pp      {\ensuremath{p}\xspace}

 \def\Ps      {\ensuremath{s}\xspace}

}

\makeatletter
\ifcase \@ptsize \relax
  \newcommand{\miniscule}{\@setfontsize\miniscule{4}{5}}
\or
  \newcommand{\miniscule}{\@setfontsize\miniscule{5}{6}}
\or
  \newcommand{\miniscule}{\@setfontsize\miniscule{5}{6}}
\fi
\makeatother

\DeclareRobustCommand{\optbar}[1]{\shortstack{{\miniscule (\rule[.5ex]{1.25em}{.18mm})}
  \\ [-.7ex] $#1$}}

\def\squark    {{\ensuremath{\Ps}}\xspace}

\def\cquark    {{\ensuremath{\Pc}}\xspace}

\def\bquark    {{\ensuremath{\Pb}}\xspace}

\def\pion   {{\ensuremath{\Ppi}}\xspace}
\def\piz    {{\ensuremath{\pion^0}}\xspace}

\def\pip    {{\ensuremath{\pion^+}}\xspace}
\def\pim    {{\ensuremath{\pion^-}}\xspace}

\def\kaon    {{\ensuremath{\PK}}\xspace}
  \def\Kbar    {{\kern 0.2em\overline{\kern -0.2em \PK}{}}\xspace}

\def\KorKbar    {\kern 0.18em\optbar{\kern -0.18em K}{}\xspace}

\def\Kp      {{\ensuremath{\kaon^+}}\xspace}
\def\Km      {{\ensuremath{\kaon^-}}\xspace}

  \def\Dbar    {{\kern 0.2em\overline{\kern -0.2em \PD}{}}\xspace}
\def\D       {{\ensuremath{\PD}}\xspace}

\def\DorDbar    {\kern 0.18em\optbar{\kern -0.18em D}{}\xspace}
\def\Dz      {{\ensuremath{\D^0}}\xspace}

\def\Dp      {{\ensuremath{\D^+}}\xspace}

\def\Dstarp  {{\ensuremath{\D^{*+}}}\xspace}

\def\Ds      {{\ensuremath{\D^+_\squark}}\xspace}

\def\Bbar    {{\ensuremath{\kern 0.18em\overline{\kern -0.18em \PB}{}}}\xspace}

\def\BorBbar    {\kern 0.18em\optbar{\kern -0.18em B}{}\xspace}

  \def\Y#1S{\ensuremath{\PUpsilon{(#1S)}}\xspace}

\def\proton      {{\ensuremath{\Pp}}\xspace}

\def\Xires       {{\ensuremath{\PXi}}\xspace}

\def\Lz          {{\ensuremath{\PLambda}}\xspace}
\def\Lbar        {{\ensuremath{\kern 0.1em\overline{\kern -0.1em\PLambda}}}\xspace}
\def\LorLbar    {\kern 0.18em\optbar{\kern -0.18em \PLambda}{}\xspace}

\def\Lc      {{\ensuremath{\Lz^+_\cquark}}\xspace}

\def\Xibz    {{\ensuremath{\Xires^0_\bquark}}\xspace}
\def\Xibm    {{\ensuremath{\Xires^-_\bquark}}\xspace}

\def\Xicp    {{\ensuremath{\Xires^+_\cquark}}\xspace}

\def\to                 {\ensuremath{\rightarrow}\xspace}

\def\AT#1     {\ensuremath{A_{\mathrm{T}}^{#1}}\xspace}           

\def\C#1      {\ensuremath{\mathcal{C}_{#1}}\xspace}                       
\def\Cp#1     {\ensuremath{\mathcal{C}_{#1}^{'}}\xspace}                    
\def\Ceff#1   {\ensuremath{\mathcal{C}_{#1}^{\mathrm{(eff)}}}\xspace}        
\def\Cpeff#1  {\ensuremath{\mathcal{C}_{#1}^{'\mathrm{(eff)}}}\xspace}       
\def\Ope#1    {\ensuremath{\mathcal{O}_{#1}}\xspace}                       
\def\Opep#1   {\ensuremath{\mathcal{O}_{#1}^{'}}\xspace}                    

\newcommand{\tev}{\ifthenelse{\boolean{inbibliography}}{\ensuremath{~T\kern -0.05em eV}\xspace}{\ensuremath{\mathrm{\,Te\kern -0.1em V}}}\xspace}
\newcommand{\gev}{\ensuremath{\mathrm{\,Ge\kern -0.1em V}}\xspace}
\newcommand{\mev}{\ensuremath{\mathrm{\,Me\kern -0.1em V}}\xspace}
\newcommand{\kev}{\ensuremath{\mathrm{\,ke\kern -0.1em V}}\xspace}
\newcommand{\ev}{\ensuremath{\mathrm{\,e\kern -0.1em V}}\xspace}
\newcommand{\gevc}{\ensuremath{{\mathrm{\,Ge\kern -0.1em V\!/}c}}\xspace}
\newcommand{\mevc}{\ensuremath{{\mathrm{\,Me\kern -0.1em V\!/}c}}\xspace}
\newcommand{\gevcc}{\ensuremath{{\mathrm{\,Ge\kern -0.1em V\!/}c^2}}\xspace}
\newcommand{\gevgevcccc}{\ensuremath{{\mathrm{\,Ge\kern -0.1em V^2\!/}c^4}}\xspace}
\newcommand{\mevcc}{\ensuremath{{\mathrm{\,Me\kern -0.1em V\!/}c^2}}\xspace}

\def\mum  {\ensuremath{{\,\upmu\rm m}}\xspace}

\def\invfb   {\ensuremath{\mbox{\,fb}^{-1}}\xspace}

\def\gsim{{~\raise.15em\hbox{$>$}\kern-.85em
          \lower.35em\hbox{$\sim$}~}\xspace}
\def\lsim{{~\raise.15em\hbox{$<$}\kern-.85em
          \lower.35em\hbox{$\sim$}~}\xspace}

\def\pt         {\mbox{$p_{\rm T}$}\xspace}

\def\evtgen     {\mbox{\textsc{EvtGen}}\xspace}

\def\geant      {\mbox{\textsc{Geant4}}\xspace}

\def\photos     {\mbox{\textsc{Photos}}\xspace}

\def\pythia     {\mbox{\textsc{Pythia}}\xspace}

\def\tell1  {TELL1\xspace}
\def\ukl1   {UKL1\xspace}

\newcommand{\eg}{\mbox{\itshape e.g.}\xspace}

\usepackage{cite} 
\usepackage{mciteplus}

\def\Xibx    {{\ensuremath{\Xires_\bquark}}\xspace}
\def\Xicx    {{\ensuremath{\Xires_\cquark}}\xspace}
\def\XibxPrime  {{\ensuremath{\Xires^{\prime}_\bquark}}\xspace}
\def\XibxStar  {{\ensuremath{\Xires^{*}_\bquark}}\xspace}
\def\XibPrimeMinus  {{\ensuremath{\Xires^{\prime -}_\bquark}}\xspace}
\def\XibPrimeZero  {{\ensuremath{\Xires^{\prime 0}_\bquark}}\xspace}
\def\XibStarMinus  {{\ensuremath{\Xires^{*-}_\bquark}}\xspace}
\def\XibStarZero  {{\ensuremath{\Xires^{*0}_\bquark}}\xspace}
\def\XibPrimeOrStarX  {{\ensuremath{\Xires^{(\prime,*)}_\bquark}}\xspace}
\def\XibPrimeOrStarMinus  {{\ensuremath{\Xires^{(\prime,*) \, -}_\bquark}}\xspace}
\def\pis  {{\ensuremath{\pi_{\mathrm{s}}^-}}\xspace}
\def\XibHigherMinus  {{\ensuremath{\Xires^{**-}_\bquark}}\xspace}
\def\XibHigherZero  {{\ensuremath{\Xires^{**0}_\bquark}}\xspace}
\newcommand{\kevcc}{\ensuremath{{\mathrm{\,ke\kern -0.1em V\!/}c^2}}\xspace}

\begin{document}

\renewcommand{\thefootnote}{\fnsymbol{footnote}}
\setcounter{footnote}{1}

\begin{titlepage}
\pagenumbering{roman}

\vspace*{-1.5cm}
\centerline{\large EUROPEAN ORGANIZATION FOR NUCLEAR RESEARCH (CERN)}
\vspace*{1.5cm}
\hspace*{-0.5cm}
\begin{tabular*}{\linewidth}{lc@{\extracolsep{\fill}}r}
\ifthenelse{\boolean{pdflatex}}
{\vspace*{-2.7cm}\mbox{\!\!\!\includegraphics[width=.14\textwidth]{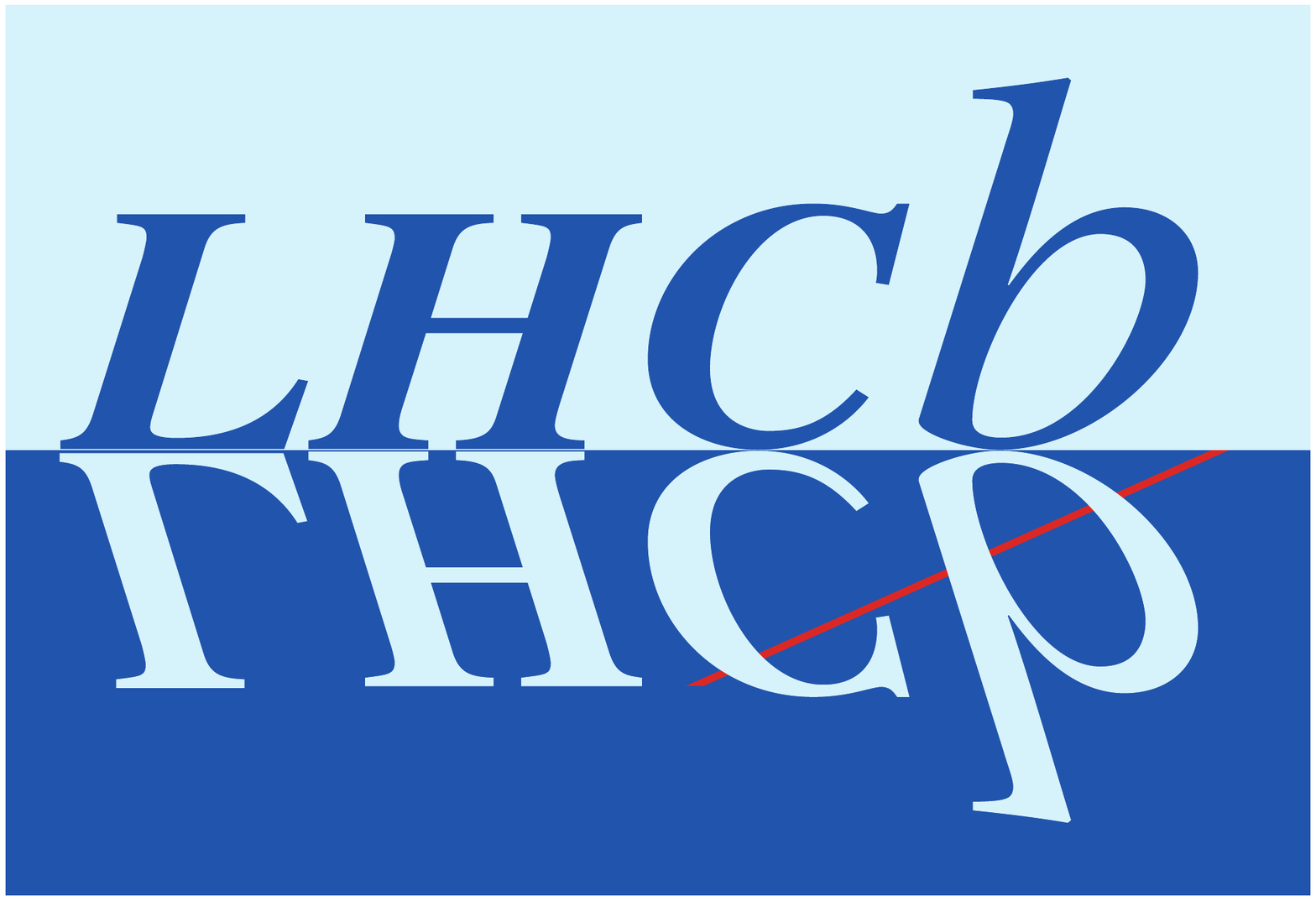}} & &}%
{\vspace*{-1.2cm}\mbox{\!\!\!\includegraphics[width=.12\textwidth]{lhcb-logo.eps}} & &}%
\\
 & & CERN-PH-EP-2014-279 \\  
 & & LHCb-PAPER-2014-061 \\  
 & & January 8, 2015 \\ 
 & & \\
\end{tabular*}

\vspace*{2.0cm} 

{\bf\boldmath\huge
\begin{center}
  Observation of two new $\Xibm$ baryon resonances
\end{center}
}

\vspace*{1.0cm} 

\begin{center}
The LHCb collaboration\footnote{Authors are listed at the end of this Letter.}
\end{center}

\vspace{\fill}

\begin{abstract}
  \noindent
  Two structures are observed close to the kinematic threshold in the $\Xibz\pim$ mass spectrum
  in a sample of proton-proton collision data,
  corresponding to an integrated luminosity of 3.0\invfb,
  recorded by the LHCb experiment.
  In the quark model, two baryonic resonances with quark content $bds$ are
  expected in this mass region: the spin-parity $J^P = \frac{1}{2}^+$ and~$J^P=\frac{3}{2}^+$
  states, denoted $\XibPrimeMinus$ and $\XibStarMinus$.
  Interpreting the structures as these resonances,
  we measure the mass differences and the width of the heavier state to be
  \begin{eqnarray*}
    m(\XibPrimeMinus) - m(\Xibz) - m(\pi^{-}) &=& 3.653 \pm 0.018 \pm 0.006 \mevcc, \\
m(\XibStarMinus) - m(\Xibz) - m(\pi^{-}) &=& 23.96 \pm 0.12 \pm 0.06 \mevcc, \\
\Gamma(\XibStarMinus) &=& 1.65 \pm 0.31 \pm 0.10 \mev

    ,
  \end{eqnarray*}  
  where the first and second uncertainties are statistical and systematic, respectively.
  The width of the lighter state is consistent with zero, and we place an
  upper limit of
  $\Gamma(\XibPrimeMinus) < 0.08$\mev at 95\% confidence level.
  Relative production rates of these states are also reported.

\end{abstract}

\vspace*{1.0cm}  

\begin{center}
  Submitted to Phys.~Rev.~Lett. \\
\end{center}

\vspace{\fill}

{\footnotesize 
\centerline{\copyright~CERN on behalf of the \lhcb collaboration, license \href{http://creativecommons.org/licenses/by/4.0/}{CC-BY-4.0}.}}
\vspace*{2mm}

\end{titlepage}


\newpage
\setcounter{page}{2}
\mbox{~}

\cleardoublepage

\renewcommand{\thefootnote}{\arabic{footnote}}
\setcounter{footnote}{0}


\pagestyle{plain} 
\setcounter{page}{1}
\pagenumbering{arabic}


In the constituent quark model~\cite{GellMann:1964nj,Zweig:1964jf},
baryonic states form multiplets according to the
symmetry of their flavor, spin, and spatial
wavefunctions. The \Xibx states form
isodoublets composed of a \Xibz ($bsu$) and a \Xibm ($bsd$) state.
Three such $\Xibx$ isodoublets that are neither orbitally nor radially excited
are expected to exist, and can be categorized by the spin $j$
of the $su$ or $sd$ diquark and the spin-parity $J^P$ of the baryon:
one with $j=0$ and $J^P = \frac{1}{2}^+$,
one with $j=1$ and $J^P = \frac{1}{2}^+$,
and one with $j=1$ and $J^P = \frac{3}{2}^+$.
This follows the same pattern as the well-known \Xicx states~\cite{PDG2014}, and 
we therefore refer to these three isodoublets as the \Xibx, the \XibxPrime, and the 
\XibxStar. The spin-antisymmetric $J^P = \frac{1}{2}^+$ state, observed
by multiple experiments~\cite{Aaltonen:2007ap,Abazov:2007am,Aaltonen:2009ny,Aaltonen:2011wd,LHCb-PAPER-2013-056,LHCb-PAPER-2014-021,Aaltonen:2014wfa,LHCb-PAPER-2012-048},
is the lightest and therefore decays through the weak interaction. The others should
decay predominantly strongly through a $P$-wave pion transition ($\XibPrimeOrStarX\to\Xibx\pi$)
if their masses are above the kinematic threshold for such a decay; otherwise
they should decay electromagnetically ($\XibPrimeOrStarX\to \Xibx \gamma$).
Observing such electromagnetic decays at hadron colliders is challenging 
due to large photon multiplicities and worse energy resolution
for low energy photons compared to charged particles. 

There are numerous predictions for the mass spectrum of these low-lying
states~\cite{Klempt:2009pi,Karliner:2008sv,Lewis:2008fu,Ebert:2005xj,Liu:2007fg,Jenkins:2007dm,Karliner:2008in,Zhang:2008pm,Wang:2010vn,Brown:2014ena,Valcarce:2008dr,Limphirat:2010zz}.
The consensus is that the isospin-averaged value of
the mass difference $m(\XibxStar) - m(\Xibx)$ is above threshold for strong decay
but that the isospin-averaged difference
$m(\XibxPrime) - m(\Xibx)$ is near the kinematic threshold. 
However, it is expected that the mass difference $m(\XibPrimeMinus) - m(\Xibz)$ is larger than
$m(\XibPrimeZero) - m(\Xibm)$ 
due to the relatively large isospin splitting between the charged and 
neutral \Xibx states. For the ground state, the measured isospin splitting of 
$m(\Xibm) - m(\Xibz) = 5.92\pm0.64$\mevcc~\cite{LHCb-PAPER-2014-048}
is in good agreement with the predicted value of $6.24 \pm 0.21$\mevcc~\cite{Karliner:2008sv}.
While the equivalent isospin splitting for the \XibxPrime and \XibxStar
states is likely to be smaller due to differences in the hyperfine mass corrections,
the mass difference $m(\XibPrimeMinus)-m(\Xibz)$ could well be
5--10\mevcc larger than $m(\XibPrimeZero)-m(\Xibm)$. It is therefore plausible
that the decay $\XibPrimeMinus \to \Xibz \pim$ is kinematically allowed, while
$\XibPrimeZero \to \Xibm\pip$ is not. This is consistent with the recent CMS 
observation~\cite{Chatrchyan:2012ni} of a single peak in the $\Xibm\pi^+$ mass spectrum, 
interpreted as the $\XibStarZero$ resonance. 
We note that $\XibPrimeZero \to \Xibz \piz$ may also be allowed
even if $\XibPrimeZero \to \Xibm \pip$ is not.

In this Letter we present the results of a study of the $\Xibz\pim$ mass
spectrum using $pp$ collision data recorded by the LHCb experiment,
corresponding to an integrated luminosity of 3.0\invfb. One third
of the data were collected at a center-of-mass energy of 7\tev and the remainder
at 8\tev. 
We observe two highly significant structures, which are interpreted as the 
\XibPrimeMinus and \XibStarMinus baryons. The properties of these new states are reported.
Charge-conjugate processes are implicitly included.


The \lhcb detector~\cite{Alves:2008zz} is a single-arm forward
spectrometer covering the \mbox{pseudorapidity} range $2<\eta <5$,
designed for the study of particles containing \bquark or \cquark
quarks. The detector includes a high-precision tracking system,
which provides a momentum measurement with precision of about 0.5\% from
2$-$100~\gevc and impact parameter resolution of approximately 20\mum for
particles with large transverse momentum (\pt). 
Ring-imaging Cherenkov detectors~\cite{LHCb-DP-2012-003}
are used to distinguish charged hadrons. Photon, electron and
hadron candidates are identified using a calorimeter system, which is followed by
detectors to identify muons~\cite{LHCb-DP-2012-002}.

The trigger~\cite{LHCb-DP-2012-004} consists of a
hardware stage, based on information from the calorimeter and muon
systems, followed by a software stage.
The software trigger requires a two-, three- or four-track
secondary vertex which is significantly displaced from
all primary $pp$ vertices (PVs) and for which the scalar \pt sum of the
charged particles is large. At least one particle should have $\pt >
1.7\gevc$ and be inconsistent with coming from any of the PVs.
A multivariate algorithm~\cite{BBDT} is used to identify
secondary vertices consistent with the decay of a \bquark hadron.

In the simulation, $pp$ collisions are generated using
\pythia~\cite{Sjostrand:2006za,*Sjostrand:2007gs} 
 with a specific \lhcb
configuration~\cite{LHCb-PROC-2010-056}.  Decays of hadrons
are described by \evtgen~\cite{Lange:2001uf}, in which final-state
radiation is generated using \photos~\cite{Golonka:2005pn}. The
interaction of the generated particles with the detector, and its
response, are implemented using the \geant
toolkit~\cite{Allison:2006ve, *Agostinelli:2002hh} as described in
Ref.~\cite{LHCb-PROC-2011-006}.


Signal candidates are reconstructed in the final state
$\Xibz \pis$, where
$\Xibz \to \Xicp \pim$ and
$\Xicp \to \proton \Km \pip$.
The first pion is denoted \pis to distinguish it
from the others.
The \Xibz decay mode is the same as that studied in~\cite{LHCb-PAPER-2014-021},
and the selection used for this analysis is heavily inspired by it and
by other LHCb studies with baryons or low-momentum pions in the final state
(\eg \cite{LHCb-PAPER-2011-023,LHCb-PAPER-2013-049}).
At each stage of the decay chain, the particles
are required to meet at a common vertex with good fit quality.
In the case of the
$\Xibz \pis$ candidate, this vertex is constrained to be consistent with
one of the PVs in the event.
Track quality requirements are applied, along with momentum and
transverse momentum requirements, to reduce combinatorial background.
Particle identification
criteria are applied to the final-state tracks to suppress background from
misidentified particles. To remove cross feed from other charm hadrons,
$\Xicp$ candidates are rejected if they are consistent with 
$\Dp \to \Kp \Km \pip$,
$\Ds \to \Kp \Km \pip$,
$\Dp \to \pip \Km \pip$, or
$\Dstarp \to \Dz(\Kp\Km)\pip$ decays. 
To reduce background formed from tracks originating at the PV,
the decay vertices of \Xicp and \Xibz candidates are
required to be significantly displaced from all PVs.

The \Xicp candidates are required to have an invariant mass within
20\mevcc of the known mass~\cite{PDG2014}, corresponding to approximately $\pm3\sigma_{\Xicp}$
where $\sigma_{\Xicp}$ is the mass resolution.
Candidate \Xibz decays are required to satisfy
$5765 < m_{\mathrm{cand}}(\Xibz) - m_{\mathrm{cand}}(\Xicp) + m_{\Xicp} < 5825$\mevcc,
where $m_{\mathrm{cand}}$ and $m_{\Xicp}$
refer to the candidate and world-average masses,
corresponding to approximately $\pm2\sigma_{\Xibz}$.
In addition, the following kinematic requirements are imposed:
$\pt(\Xicp) > 1$\gevc,
$\pt(\Xibz) > 2$\gevc,
$\pt(\Xibz \pis) > 2.5$\gevc, and
$\pt(\pis) > 0.15$\gevc.
Defining $\delta m \equiv m_{\mathrm{cand}}(\Xibz \pis) - m_{\mathrm{cand}}(\Xibz) - m_{\pim}$,
the region of consideration is $\delta m < 45$\mevcc.
There are on average 1.15
candidates retained in this region per event. Such multiple candidates 
are due almost entirely to cases where
the same \Xibz candidate is combined with different
\pis candidates from the same PV.
All $\Xibz\pi_s^-$ candidates are kept.

\begin{figure}
  \begin{center}
    \includegraphics[width = 0.98\columnwidth]{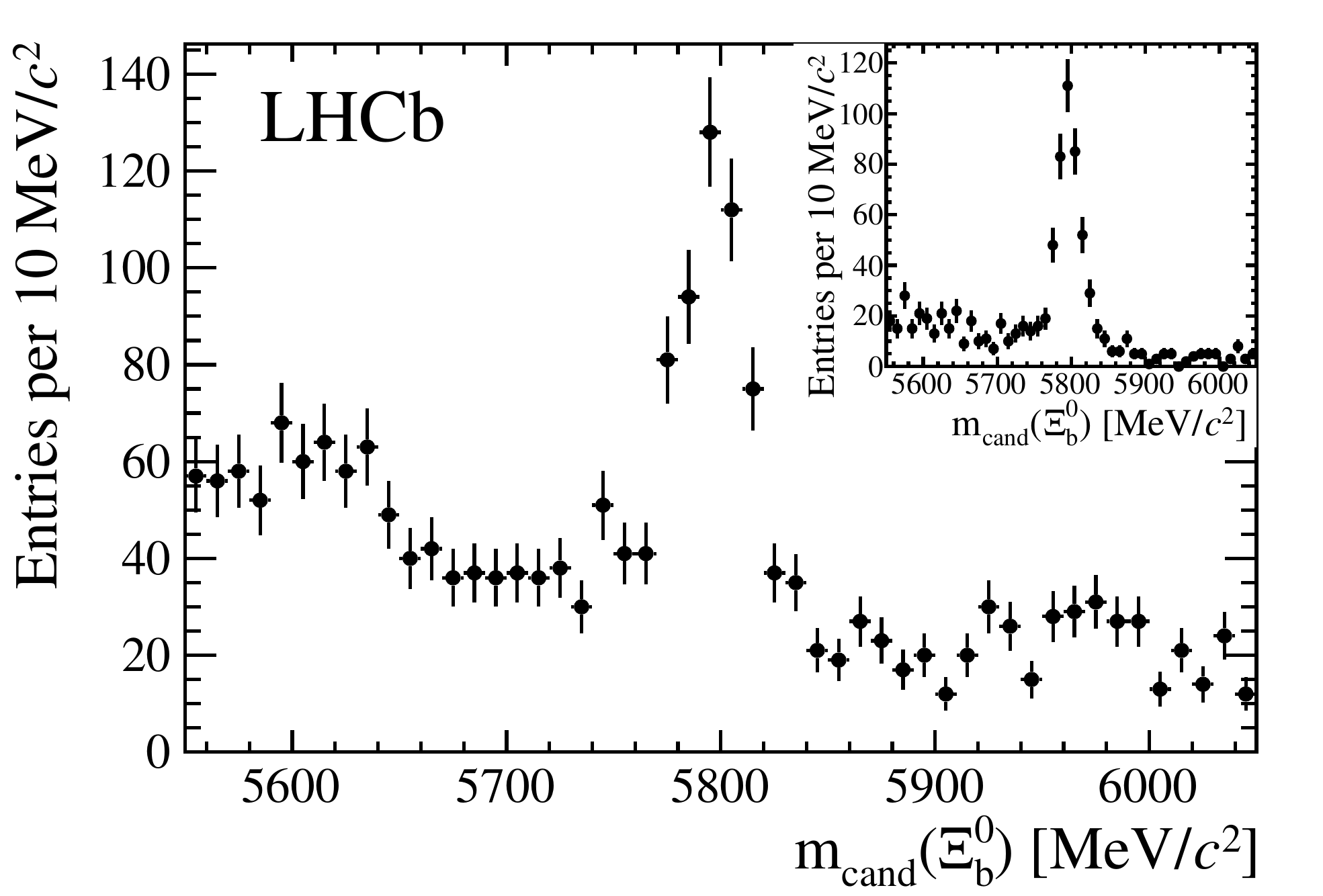}
  \end{center}
  \caption{
    Distribution of $m_{\mathrm{cand}}(\Xibz)$ for $\Xibz \pis$ candidates
    passing the full selection apart from the $m_{\mathrm{cand}}(\Xibz)$ requirement.
    Inset: The subset of candidates that lie in the $\delta m$ signal regions of
    $3.0 < \delta m < 4.2$\mevcc and $21 < \delta m < 27$\mevcc.
  }
  \label{fig:massXibZero}
\end{figure}

\begin{figure}
  \begin{center}
    \includegraphics[width = 0.98\columnwidth]{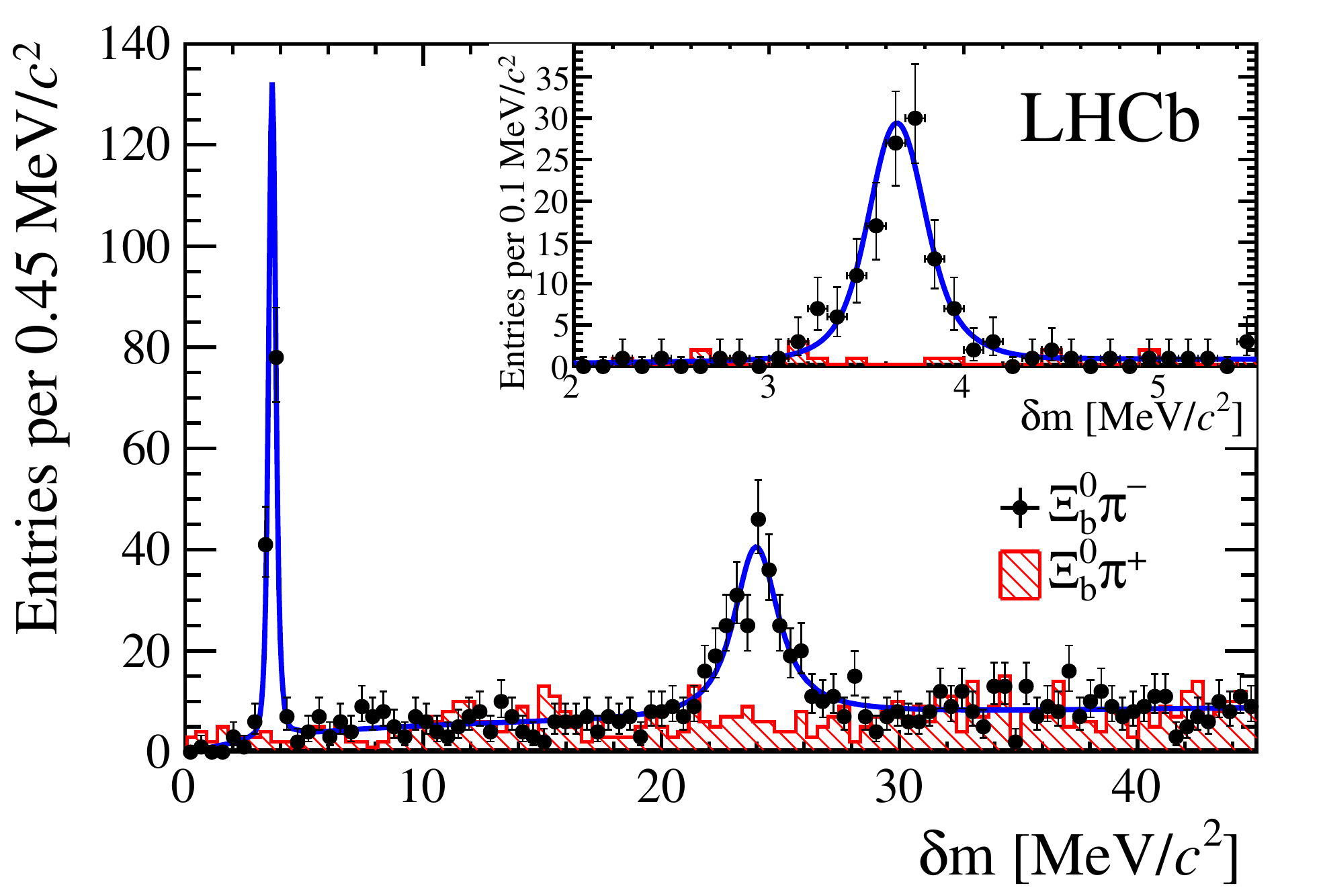}
  \end{center}
  \caption{
    Distribution of the mass difference, $\delta m$, 
    for $\Xibz \pis$ candidates in data.
    The points with error bars show
    right-sign candidates in the \Xibz mass signal region,
    and the hatched histogram shows
    wrong-sign candidates with the same selection.
    The curve shows the nominal fit to the right-sign candidates.
    Inset: detail of the region $2.0$--$5.5$\mevcc.
  }
  \label{fig:merged}
\end{figure}

The $m_{\mathrm{cand}}(\Xibz)$ projection of the $\Xibz \pis$ candidates
passing the full selection apart from the $m_{\mathrm{cand}}(\Xibz)$ requirement,
but including the $\delta m$ requirement, 
is shown in Fig.~\ref{fig:massXibZero}.
Control samples, notably wrong-sign combinations $\Xibz\pip$,
are also used to study backgrounds.
The $\delta m$ spectra for the signal and the wrong-sign sample
are shown in Fig.~\ref{fig:merged}.
Two peaks are clearly visible, a narrow one at $\delta m \approx 3.7$\mevcc
and a broader one at $\delta m \approx 24$\mevcc.
No structure is observed in the wrong-sign sample, nor in
studies of the \Xibz mass sidebands.

Accurate determination of the masses, widths, and signal yields of these two states
requires knowledge of the signal shapes, and in particular, the mass resolution of 
the two peaks. These are obtained from large samples of simulated decays with
$\delta m$ values of 3.69\mevcc and 23.69\mevcc, corresponding to the two peaks. The natural widths,
$\Gamma$, are set to negligible values so that the width measured in simulation is due
entirely to the mass resolution. The resolution function 
is parameterized as the sum of three Gaussian distributions with independent mean values.
Separate sets of parameters are determined for the two peaks.
An indication of the scale of the resolution is given by the weighted averages of
the three Gaussian widths, which are 0.21\mevcc and 0.54\mevcc for the lower- and higher-mass peaks.
In the nominal fits to data,
the parameters of the three Gaussian distributions are kept fixed
to the values obtained from simulation,
given in the Supplementary Material~\cite{bib:SupplementaryMaterial}.
Small corrections, obtained from simulation, are applied to the masses
to account for offsets in the resolution functions.
The combinatorial background is modeled by a threshold function of the form
 \begin{displaymath}
   f(\delta m) = \left( 1 - e^{-\delta m / C} \right) \, (\delta m)^{A},
   \label{eq:bkg}
 \end{displaymath}
where $A$ and $C$ are freely varying parameters determined in the fit to the data.

The masses, widths and yields of the two peaks are determined from an
unbinned maximum likelihood fit to the $\delta m$ spectrum.
In an initial fit, each peak is described using a
$P$-wave relativistic Breit-Wigner (RBW) line shape~\cite{Jackson:1964zd}
with a Blatt-Weisskopf barrier factor~\cite{Blatt:1952},
convolved with the resolution
function obtained from simulation. The fitted width of the lower-mass
peak is found to be consistent with zero
and consequently its width is set to zero in the nominal fit,
shown in Fig.~\ref{fig:merged}.
The fitted yields in the lower- and higher-mass peaks are
$121 \pm 12$ and $237 \pm 24$ events, 
with statistical significances in excess of 10$\sigma$.
The nonzero value of the natural width of the higher-mass peak is also highly significant:
the change in likelihood when the width is fixed to zero corresponds to a
$p$-value of $4 \times 10^{-14}$ using Wilks's theorem~\cite{Wilks:1938dza}.

An upper limit on the natural width of the lower-mass peak is set using ensembles
of pseudoexperiments with the same parameters as in data,
but with natural widths ranging from 0.01 to 0.12\mev. The upper limit is
taken to be the value of $\Gamma$ for which a width equal to or greater than that obtained in 
data is observed in 95\% of the pseudoexperiments.
The resulting upper limit is $\Gamma(\XibPrimeMinus) < 0.08$\mev at 95\% confidence level (CL).

A number of cross checks are performed to ensure the robustness of the
measured masses and natural widths of these states and to assess
systematic uncertainties. These include
  changing the assumed angular momentum (spin 0,\,2) and radial parameter (1--5\gev$^{-1}$) of the RBW and barrier factor;
  inflating the widths of the resolution functions by a fixed factor of $1.1$, the
  value found in a large $\Dstarp\to\Dz\pi$ data sample~\cite{LHCb-CONF-2013-003};
  inflating the widths of the resolution functions by a common factor floated in the fit (with $1.03 \pm 0.11$ obtained);
  using a symmetric resolution function;
  using a non-relativistic BW for the higher-mass peak;
  using a different background function;
  varying the fit range;
  checking the effect of finite sample size and of the variation of mass resolution with particle mass;
  keeping only one candidate in each event;
  imposing additional trigger requirements;
  separating the data by charge and LHCb magnet polarity;
  and fitting the wrong-sign sample.
Where appropriate, systematic uncertainties are assigned based on the differences between
the nominal results and those obtained in these tests.
The calibration of the momentum scale~\cite{LHCb-PAPER-2012-048,LHCB-PAPER-2013-011} is validated by measuring $m(\Dstarp)-m(\Dz)$
in a large sample of $\Dstarp$, $\Dz\to\Km\Kp$ decays~\cite{LHCb-CONF-2013-003}. The mass difference agrees
with a recent BaBar measurement\cite{Lees:2013uxa,*Lees:2013zna} within 6\kevcc,
corresponding to $1.3\sigma$ when including the mass scale uncertainty for that decay.
The uncertainties are summarized in
Table~\ref{tab:sys}. Taking these into account, we obtain
 \begin{eqnarray*}
   \delta m(\XibPrimeMinus) &=& 3.653 \pm 0.018 \pm 0.006 \mevcc, \\
\delta m(\XibStarMinus) &=& 23.96 \pm 0.12 \pm 0.06 \mevcc, \\
\Gamma(\XibStarMinus) &=& 1.65 \pm 0.31 \pm 0.10 \mev
 ,\\
   \Gamma(\XibPrimeMinus) &<& 0.08\,\mev~{\mbox{\rm at~95\%~CL}}.
 \end{eqnarray*}
Combining these with the measurement of
$m(\Xibz) = 5791.80 \pm 0.50$\mevcc obtained previously at LHCb~\cite{LHCb-PAPER-2014-021},
the masses of these states are found to be
 \begin{eqnarray*}
   m(\XibPrimeMinus) &=& 5935.02 \pm 0.02 \pm 0.01 \pm 0.50 \mevcc, \\
m(\XibStarMinus)  &=& 5955.33 \pm 0.12 \pm 0.06 \pm 0.50 \mevcc

   ,
 \end{eqnarray*}
where the uncertainties are statistical, systematic, and
due to the $m(\Xibz)$ measurement, respectively.

\begin{table}
  \caption{
    Systematic uncertainties, in units of \mevcc (masses) and \mev (width).
    The statistical uncertainties are also shown for comparison.
  }
   \begin{center}
     \begin{tabular}{lccc}
       Source & $\delta m(\XibxPrime)$ & $\delta m(\XibxStar)$ & $\Gamma(\XibxStar)$ \\ \hline
       Simulated sample size & 0.002 & 0.005 &  \\
Multiple candidates & 0.004 & 0.048 & 0.055 \\
Resolution model & 0.002 & 0.003 & 0.070 \\
Background description & 0.001 & 0.003 & 0.019 \\
Momentum scale & 0.003 & 0.014 & 0.003 \\
RBW spin and radial parameter & 0.000 & 0.023 & 0.028 \\
\hline Sum in quadrature & 0.006 & 0.055 & 0.095 \\
Statistical uncertainty & 0.018 & 0.119 & 0.311
     \end{tabular}
   \end{center}
  \label{tab:sys}
\end{table}

Helicity angle~\cite{Richman:1984gh} distributions may be used
to distinguish between spin hypotheses for resonances.
We consider the decay sequence $\XibPrimeOrStarMinus \to \Xibz \pim$, $\Xibz \to \Xicp \pim$,
where the \XibPrimeOrStarMinus has spin $J$ and the
\Xibz, \Xicp, and \pim have spin-parity 
$\frac{1}{2}^+$, $\frac{1}{2}^+$, and $0^-$, respectively, which is analogous to 
the scenario considered in Ref.~\cite{Abe:2006rz}. Defining $\theta_h$ as the
angle between the three-momentum of the \Xibz in the
\XibPrimeOrStarMinus rest frame and the three-momentum of the
\Xicp in the \Xibz rest frame,
the $\cos\theta_h$ distribution is 
a polynomial of order $(2J-1)$.
For $J=\frac{1}{2}$ this would yield a flat distribution,
and hence a nonuniform distribution would imply $J>\frac{1}{2}$.
The converse does not follow, however: 
a higher-spin resonance that is unpolarized will lead to a flat distribution.
For each of the two peaks, the background-subtracted,
efficiency-corrected $\cos\theta_h$ distributions are
studied. Both are found to be consistent with
flat distributions.
When fitted with a function of the form
$f(\cos\theta_h) = \left[ a + 3(1-a)\cos^2\theta_h \right]/2$,
the fitted values of $a$ are $0.89 \pm 0.11$ and $0.88 \pm 0.11$,
and the quality of the fits does not improve significantly.
Thus, the available data are consistent with the quark model expectations that
the lower-mass peak corresponds to a $J=\frac{1}{2}$ state
and the higher one to a $J=\frac{3}{2}$ state (if unpolarized or weakly polarized),
but other values of $J$ are not excluded.

We measure the production rates of the two signals relative to 
that of the \Xibz state, selected inclusively and 
passing the same \Xibz selection criteria as the signal sample.
To remain within the bandwidth restrictions of the offline data reduction process,
10\% of the candidates in the normalization mode are randomly selected and
retained for use in this analysis.
To ensure that the efficiencies are well understood,
we use only the subset of events in which one or more of the \Xibz decay products 
is consistent with activating the hardware trigger in the calorimeter.

For this subsample of events, the fitted yields are
$93 \pm 10$ for the lower-mass $\Xibz \pis$ state,
$166 \pm 20$ for the higher-mass $\Xibz \pis$ state,
and $162 \pm 15$ for the \Xibz normalization sample.
The efficiency ratios are determined with simulated decays, applying the same
trigger, reconstruction, and selection procedures that are used for the data.
Systematic uncertainties (and, where appropriate, corrections) are assigned for
those sources that do not cancel in the efficiency ratios. These uncertainties include
  the modeling of the \Xibx momentum spectra,
  the \pis reconstruction efficiency~\cite{LHCb-DP-2014-002},
  the fit method,
  and the efficiency of those selection criteria that are applied to the
  $\Xibz \pis$ candidates but not to the \Xibz normalization mode.
Combining the 7\tev and 8\tev data samples,
the results obtained are
 \begin{eqnarray*}
   \frac{
     \sigma(pp \to \XibPrimeMinus X) 
     \mathcal{B}(\XibPrimeMinus \to \Xibz \pim)
   }{
     \sigma(pp \to \Xibz X)
   }
   &=& 0.118 \pm 0.017 \pm 0.007, \\
   \frac{ 
     \sigma(pp \to \XibStarMinus X) 
     \mathcal{B}(\XibStarMinus \to \Xibz \pim) 
   }{
    \sigma(pp \to \Xibz X) 
   }
   &=& 0.207 \pm 0.032 \pm 0.015, \\
   \frac{
     \sigma(pp \to \XibStarMinus X) 
     \mathcal{B}(\XibStarMinus \to \Xibz \pim) }{ \sigma(pp \to \XibPrimeMinus X) \mathcal{B}(\XibPrimeMinus \to \Xibz \pim)
   }
   &=& 1.74 \pm 0.30 \pm 0.12 ,
 \end{eqnarray*}
where
the first and second uncertainties are statistical and systematic, respectively,
$\sigma$ denotes a cross-section measured within the LHCb acceptance
and extrapolated to the full kinematic range with \pythia,
$\mathcal{B}$ represents a branching fraction, and
$X$ refers to the rest of the event.
Given that isospin partner modes
$\XibPrimeZero \to \Xibz \piz$ and $\XibStarZero \to \Xibz \piz$
are also expected, these results imply that a large fraction of \Xibz baryons
in the forward region are produced in the decays of \Xibx resonances.

As a further check, the $\Xibz \pis$ mass spectrum is studied with
additional \Xibz decay modes.
Significant peaks are seen with the mode 
$\Xibz \to \Lc (\proton \Km \pip) \Km \pip \pim$
for both \XibPrimeMinus $(6.4\sigma)$ and \XibStarMinus $(4.7\sigma)$.
The peaks are also seen with reduced significance in other \Xibz final states:
$4\sigma$ for \XibPrimeMinus and $2\sigma$ for \XibStarMinus in $\Xibz \to \Dz (\Km \pip) p \Km$; and
$3\sigma$ for \XibPrimeMinus and $3\sigma$ for \XibStarMinus in $\Xibz \to \Dp (\Km \pip \pip) p \Km \pim$.
The modes $\Xibz \to \Lc (\proton \Km \pip) \Km \pip \pim$ and $\Xibz \to \Dp (\Km \pip \pip) p \Km \pim$
have not been observed before, and are being studied in separate analyses.

With a specific configuration of other excited \Xibx states, it is possible
to produce a narrow peak in the $\Xibz \pim$ mass spectrum
that is not due to a \XibPrimeMinus resonance.
This can arise from the decay chain
$\XibHigherMinus \to \XibPrimeZero \pim, ~ \XibPrimeZero \to \Xibz \piz$,
where the \XibHigherMinus is the $L=1$, $J^P = \frac{1}{2}^-$ state analogous to the $\Xicx(2790)$.
If both decays are close to threshold, the particles produced will be kinematically correlated such that
combining the \Xibz daughter with the \pim from the \XibHigherMinus
would produce a structure in the $m(\Xibz \pim)$ spectrum.
In general such a structure would be broader than that seen in Fig.~\ref{fig:merged}
and would be accompanied
by a similar peak in the wrong-sign $\Xibz \pip$ spectrum from the 
isospin-partner decay, $\XibHigherZero \to \XibPrimeMinus \pip, ~ \XibPrimeMinus \to \Xibz \pim$.
However, if a number of conditions are fulfilled, including the \XibHigherMinus and \XibPrimeZero states being  $279.0\pm0.5$ and 
$135.8\pm0.5$\mevcc heavier than the \Xibz ground state, respectively,
it is possible to circumvent these constraints.
This would also require that the production rate of the
$L=1$ state be comparable to that of the $L=0$, $J^P=\frac{3}{2}^+$ state.
Although this scenario is contrived, it cannot be excluded at present.

In conclusion, two structures are observed with high significance in the
$\Xibz \pim$ mass spectrum with mass differences above threshold of 
$\delta m=3.653\pm0.018\pm0.006$\mevcc and $23.96\pm0.12\pm0.06$\mevcc.
These values are in general agreement with quark model expectations for the
$J^P=\frac{1}{2}^+$ \XibPrimeMinus and $J^P=\frac{3}{2}^+$ \XibStarMinus states.
Their natural widths are measured to be $\Gamma(\XibPrimeMinus)<0.08$\mev
at 95\% CL and $\Gamma(\XibStarMinus)=1.65\pm0.31\pm0.10$\mev.
The observed angular distributions in the decays of these states are 
consistent with the spins expected in the quark model,
but other $J$ values are not excluded.
The relative production rates are also measured.


We express our gratitude to our colleagues in the CERN
accelerator departments for the excellent performance of the LHC. We
thank the technical and administrative staff at the LHCb
institutes. We acknowledge support from CERN and from the national
agencies: CAPES, CNPq, FAPERJ and FINEP (Brazil); NSFC (China);
CNRS/IN2P3 (France); BMBF, DFG, HGF and MPG (Germany); INFN (Italy); 
FOM and NWO (The Netherlands); MNiSW and NCN (Poland); MEN/IFA (Romania); 
MinES and FANO (Russia); MinECo (Spain); SNSF and SER (Switzerland); 
NASU (Ukraine); STFC (United Kingdom); NSF (USA).
The Tier1 computing centres are supported by IN2P3 (France), KIT and BMBF 
(Germany), INFN (Italy), NWO and SURF (The Netherlands), PIC (Spain), GridPP 
(United Kingdom).
We are indebted to the communities behind the multiple open 
source software packages on which we depend. We are also thankful for the 
computing resources and the access to software R\&D tools provided by Yandex LLC (Russia).
Individual groups or members have received support from 
EPLANET, Marie Sk\l{}odowska-Curie Actions and ERC (European Union), 
Conseil g\'{e}n\'{e}ral de Haute-Savoie, Labex ENIGMASS and OCEVU, 
R\'{e}gion Auvergne (France), RFBR (Russia), XuntaGal and GENCAT (Spain), Royal Society and Royal
Commission for the Exhibition of 1851 (United Kingdom).

\clearpage
\addcontentsline{toc}{section}{References}
\setboolean{inbibliography}{true}
\bibliographystyle{LHCb}
\bibliography{main,LHCb-PAPER,LHCb-CONF,LHCb-DP,LHCb-TDR,papers}

\clearpage


\newpage
\centerline{\large\bf LHCb collaboration}
\begin{flushleft}
\small
R.~Aaij$^{41}$, 
B.~Adeva$^{37}$, 
M.~Adinolfi$^{46}$, 
A.~Affolder$^{52}$, 
Z.~Ajaltouni$^{5}$, 
S.~Akar$^{6}$, 
J.~Albrecht$^{9}$, 
F.~Alessio$^{38}$, 
M.~Alexander$^{51}$, 
S.~Ali$^{41}$
% ...

{\footnotesize \it
% ...
$ ^{5}$Clermont Universit\'{e}, Universit\'{e} Blaise Pascal, CNRS/IN2P3, LPC, Clermont-Ferrand, France\\
$ ^{6}$CPPM, Aix-Marseille Universit\'{e}, CNRS/IN2P3, Marseille, France\\
% ...
$ ^{9}$Fakult\"{a}t Physik, Technische Universit\"{a}t Dortmund, Dortmund, Germany\\
% ...
$ ^{37}$Universidad de Santiago de Compostela, Santiago de Compostela, Spain\\
$ ^{38}$European Organization for Nuclear Research (CERN), Geneva, Switzerland\\
% ...
$ ^{41}$Nikhef National Institute for Subatomic Physics, Amsterdam, The Netherlands\\
% ...
$ ^{46}$H.H. Wills Physics Laboratory, University of Bristol, Bristol, United Kingdom\\
% ...
$ ^{51}$School of Physics and Astronomy, University of Glasgow, Glasgow, United Kingdom\\
$ ^{52}$Oliver Lodge Laboratory, University of Liverpool, Liverpool, United Kingdom\\
% ...
}
\end{flushleft}

\end{document}